\catcode`\@=11  
\def\answ{o}    
\def\twocol{t }      

\documentstyle[aps,prd,tighten,epsf]{revtex}

\begin{document}
\draft
\if\answ\twocol
      \twocolumn[\hsize\textwidth\columnwidth\hsize\csname
             @twocolumnfalse\endcsname
\fi

\title{Late-time evolution of a self-interacting
 scalar field in
the spacetime of dilaton black hole}

\author{ Rafa{\l} Moderski}
\address{
Nicolaus Copernicus Astronomical Center \protect \\
Polish Academy of Sciences \protect \\
00-716 Warsaw, Bartycka 18, Poland \protect \\
moderski@camk.edu.pl }

\author{Marek Rogatko}
\address{Technical University of Lublin \protect \\
20-618 Lublin, Nadbystrzycka 40, Poland \protect \\
rogat@tytan.umcs.lublin.pl \protect \\
rogat@akropolis.pol.lublin.pl}
\date{\today}
\maketitle

\if\answ\twocol
     \widetext
\fi
\begin{abstract}
We investigate the late-time tails of self-interacting (massive)
scalar fields in the spacetime of dilaton black hole. Following the
{\it no hair} theorem we examine the mechanism by which
self-interacting scalar hair decay. We revealed that the intermediate
asymptotic behavior of the considered field perturbations is dominated
by an oscillatory inverse power-law decaying tail.  The numerical
simulations showed that at the very late-time massive self-interacting
scalar hair decayed slower than any power law.

\end{abstract}

\if\answ\twocol
     \vskip 2pc]
     \narrowtext
\fi
\newcommand{\be}{\begin{equation}}
\newcommand{\ee}{\end{equation}}
\newcommand{\ben}{\begin{eqnarray}}
\newcommand{\een}{\end{eqnarray}}

\newcommand{\la}{{\lambda}}
\newcommand{\Om}{{\Omega}}
\newcommand{\ta}{{\tilde a}}
\newcommand{\bg}{{\bar g}}
\newcommand{\bh}{{\bar h}}
\newcommand{\si}{{\sigma}}
\newcommand{\th}{{\theta}}
\newcommand{\C}{{\cal C}}
\newcommand{\D}{{\cal D}}
\newcommand{\cA}{{\cal A}}
\newcommand{\cT}{{\cal T}}
\newcommand{\cO}{{\cal O}}
\newcommand{\eeo}{\cO ({1 \over E})}
\newcommand{\G}{{\cal G}}
\newcommand{\cL}{{\cal L}}
\newcommand{\T}{{\cal T}}
\newcommand{\M}{{\cal M}}

\newcommand{\p}{\partial}
\newcommand{\na}{\nabla}
\newcommand{\ssum}{\sum\limits_{i = 1}^3}
\newcommand{\dssum}{\sum\limits_{i = 1}^2}
\newcommand{\tal}{{\tilde \alpha}}

\newcommand{\tp}{{\tilde \phi}}
\newcommand{\tPhi}{\tilde \Phi}
\newcommand{\tpsi}{\tilde \psi}
\newcommand{\tim}{{\tilde \mu}}
\newcommand{\tr}{{\tilde \rho}}
\newcommand{\tir}{{\tilde r}}
\newcommand{\rp}{r_{+}}
\newcommand{\hr}{{\hat r}}
\newcommand{\rv}{{r_{v}}}
\newcommand{\dr}{{d \over d \hr}}
\newcommand{\dR}{{d \over d R}}

\newcommand{\hhf}{{\hat \phi}}
\newcommand{\hhM}{{\hat M}}
\newcommand{\hhQ}{{\hat Q}}
\newcommand{\hht}{{\hat t}}
\newcommand{\hhr}{{\hat r}}
\newcommand{\hhS}{{\hat \Sigma}}
\newcommand{\hhD}{{\hat \Delta}}
\newcommand{\hhm}{{\hat \mu}}
\newcommand{\hro}{{\hat \rho}}
\newcommand{\hhz}{{\hat z}}

\newcommand{\tD}{{\tilde D}}
\newcommand{\tB}{{\tilde B}}
\newcommand{\tV}{{\tilde V}}
\newcommand{\hT}{\hat T}
\newcommand{\tF}{\tilde F}
\newcommand{\tT}{\tilde T}
\newcommand{\hC}{\hat C}
\newcommand{\ep}{\epsilon}
\newcommand{\bep}{\bar \epsilon}
\newcommand{\ppp}{\varphi}
\newcommand{\Ga}{\Gamma}
\newcommand{\ga}{\gamma}
\newcommand{\hth}{\hat \theta}
\bigskip

\baselineskip=18pt
\par
\section{Introduction}
The late-time evolution of various fields outside a collapsing body
plays an important role in major aspects of black hole physics, as the
{\it no-hair} theorem \cite{bh} and the mass inflation scenario
\cite{mi}. Wheeler coined the metaphoric dictum {\it black holes have
no hair} which means that regardless of the specific details of the
collapse or structure and properties of the collapsing body a
stationary black hole characterized by its mass, charge and angular
momentum emerged in the resultant process. Therefore it is interesting
to study the mechanism responsible for the decay of black hole hair.\\
The neutral external perturbations were first studied in \cite{pri}
and it was found that the late-time behavior is dominated by the
factor $t^{-(2l + 3)}$, for each multipole moment $l$. The neutral
perturbations along null infinity and along the future event horizon
were studied in Ref.\cite{gun}. Their decays were governed by the
power laws $u^{-(l + 2)}$ and $v^{-(l + 3)}$, where $u$ and $v$ were
the outgoing Eddington-Filkenstein (ED) and ingoing ED coordinates.
Bicak \cite{bic} considered the scalar perturbations on
Reissner-Nordtr\"om (RN) background and found that for $\mid Q \mid <
M$ the relation $t^{-(2l + 2)}$, while for $\mid Q \mid = M$ the
late-time behavior at fixed $r$ was governed by $t^{-(l + 2)}$.
\par
Hod and Piran \cite{pir1}-\cite{pir3} studied the late time behavior
of a charged massless scalar field in RN spacetime. Among all the
conclusion was that a charged hair decayed slower than a neutral
one.
\par
The problem of late-time tails in gravitational collapse of a
self-interacting (SI) fields in the background of Schwarzschild
solution was reported by Burko \cite{bur} and in RN spacetime, at
intermediate late-time, was considered in Ref.\cite{mm}. At
intermediate late-time for small mass $m$ the decay was dominated by
the oscillatory inverse power tails $t^{-(l +3/2)} \sin (m t)$. This
analytic prediction was verified at intermediate times, where $mM \le
mt \le 1/(mM)^2$. In Ref.\cite{ja} the nearly extreme RN spacetime was
considered and it was found analytically that the inverse power-law
behavior of the dominant asymptotic tail is of the form $t^{-5/6} \sin
(m t)$, independent of $l$. The asymptotic tail behavior of SI scalar
field was also studied in Schwarzschild spacetime \cite{ja1}. The
oscillatory tail of scalar field has the decay rate of $t^{-5/6}$ at
asymptotically late time.
\par
Nowadays, it seems that the superstring theories are the most
promising candidates for a consistent quantum theory of
gravitation. In Ref.\cite{rog} the asymptotic evolution of a massless
charged scalar field in the background of dilaton black hole was
studied. Using both numerical and analytical methods the inverse
power-law relaxation was revealed. It turned out that the charged hair
decayed slower than a neutral one.
\par
In our work we shall generalize our previous considerations \cite{rog}
and discuss the SI scalar field behavior in the spacetime of dilaton
black hole.  In Sec.II we gave the analytic arguments concerning the
intermediate and the very late-time behavior of SI scalar field in the
background of the considered black hole.  Then, in Sec.III we treated
the problem numerically. Sec.IV will be devoted to a summary and
conclusions.

\section{The Einstein-Maxwell-dilaton Equations}
\label{sec1}
In this section we shall study the evolution of SI (massive) scalar
field $\tpsi$ around a fixed background of electrically charged
dilaton black hole. The wave equation for the field is given by
\be
\na_{\mu} \na^{\mu} \tpsi 
- m^2 \tpsi^2 = 0.
\ee
where $m$ is assumed to be real.
\par
The metric of the external gravitational field will be given by the
static, spherically symmetric solution of Eqs. of motion derived from
the low-energy string action (see e.g.\cite{dbh}). The action has the
form as follows:
\be
S = \int d^4 x \left [
R - 2 (\na \phi)^2 - e^{-2 \phi} F^2 \right ],
\ee
where $\phi$ is the dilatonic field and $F_{\alpha \beta} = 2\na_{[
\alpha} A_{\beta ]}$.
The metric of electrically charged dilaton black hole
\cite{dbh} implies
\be
ds^2 = - \left ( 1 - {2 M \over r} \right )dt^2 +
{dr^2 \over \left ( 1 - {2 M \over r} \right )} +
r \left ( r - {Q^2 \over M} \right ) (d\theta^2 + \sin \theta d \phi^2 ).
\label{met}
\ee
The event horizon is located at $r_{+} = 2 M$. For the case of $r_{-}
= {Q^2 \over M}$ we have another singularity but it can be ignored
because of the fact that $r_{-} < r_{+}$. The dilaton field is given
by $e^{2 \phi} = e^{- 2 \phi_{0}} \left ( 1 - {r_{-} \over r} \right
)$, where $\phi_{0}$ is the dilaton's value at $r \rightarrow
\infty$. The mass $M$ and the charge $Q$ are related by the relation
$Q^2 = {r_{+}r_{-} \over 2} e^{2 \phi_{0}}$.\\
Defining the tortoise coordinate $y$, as $dy = {dr \over \left ( 1 -
{2 M \over r} \right ) }$ one can rewrite (\ref{met}) in the form
\be
ds^2 = \left ( 1 - {2 M \over r} \right ) \left [
- dt^2 + dy^2 \right ] + r \left ( r - {Q^2 \over M} \right ) 
(d\theta^2 + \sin \theta d \phi^2 ).
\ee
For the spherical background each of the multipole of perturbation
field evolves separetly so for the scalar field in the form $\tpsi =
\sum_{l,m} \psi_{m}^{l}(t, r) Y_{l}^{m}(\theta, \phi)/R(r)$, one has
the following equations of motion for each multipole moment
\be
\psi_{,tt} - \psi_{,yy} + V \psi = 0,
\label{mo}
\ee
where
\be
V = \left [
{l(l + 1) \over R^2} + {R'' \over R} \left ( 1 - {2 M \over r} \right )
+ {2 M R' \over r^2 R} + m^2 
\right ] \left ( 1 - {2 M \over r} \right ).
\ee
By $R$ we denoted $R = \sqrt{r \left ( r - {Q^2 \over M} \right )}$
and $'$ is the derivative with respect to the $r$-coordinate.
\par
In order to analyze the time evolution of SI field in the background
of dilaton black hole we shall use the spectral decomposition method
\cite{lea}.
The time evolution of SI scalar field is given by
\be
\psi(y, t) = \int dy' \bigg[ G(y, y';t) \psi_{t}(y', 0) +
G_{t}(y, y';t) \psi(y', 0) \bigg],
\ee
for $t > 0$, where   the Green's function  $ G(y, y';t)$ is defined by
\be
\bigg[ {\p^2 \over \p t^2} - {\p^2 \over \p y^2 } + V \bigg]
G(y, y';t)
= \delta(t) \delta(y - y').
\label{green}
\ee
Our main task will be to find the black hole Green function.  The
first step in finding $G(y, y';t)$ consists of reducing (\ref{green})
to an ordinary differential equation.  To do it one can use the
Fourier transform \cite{and}
\be
\tilde  
G(y, y';\omega) = \int_{0^{-}}^{\infty} dt  G(y, y';t) e^{i \omega t}.
\ee
This transform is well defined if $Im~ \omega \ge 0$, while the
corresponding inverse transform has the form
\be
G(y, y';t) = {1 \over 2 \pi} \int_{- \infty + i \ep}^{\infty + i \ep}
d \omega
\tilde G(y, y';\omega) e^{- i \omega t},
\ee
for some positive number $\ep$.

The Fourier's component of the Green's function $\tilde G(y,
y';\omega)$ can be expressed in terms of two linearly independent
solutions for homogeneous equation, namely
\be
\bigg(
{d^2 \over dy^2} + \omega^2 - V \bigg) \psi_{i} = 0, \qquad i = 1, 2.
\label{wav}
\ee
The boundary conditions for $\psi_{i}$ are described by purely ingoing
waves crossing the outer horizon $H_{+}$ of the dilaton black hole
$\psi_{1} \simeq e^{- i \omega y}$ as $y \rightarrow - \infty$ while
$\psi_{2}$ should be damped expotentially at $i_{+}$, namely $\psi_{2}
\simeq e^{- \sqrt{\omega^2 - m^2}y}$ at $y \rightarrow \infty$.
\par
In order to find $\psi_{i}$ we consider the wave Eq.(\ref{wav}) of SI
scalar field and transform it in such a way that \cite{pir2} the
Coulomb and Newtonian $(1/r)$ potentials will dominate at large
distances. One can introduce an auxiliary variable $\xi$ in such a way
that $\xi = A(r) \psi$, where $A(r)^2 = 1 - {2M \over r}$.  So in
terms of the variable $\xi$ Eq.(\ref{wav}) can be rewritten as
follows:
\be
{d^2 \xi \over d r^2} - {A_{,rr} \over A} \xi + {1 \over A^4}
\bigg[ \omega^2 - A^2 \bigg(
{l(l + 1) \over R^2} + {R'' \over R} A(r)^2
+ {2 M R' \over r^2 R} + m^2 
\bigg) \bigg] \xi = 0.
\label{wav1}
\ee
Now we expand Eq.(\ref{wav1}) in power series of ${M \over r}$ and ${Q
\over r}$, neglecting terms of order $ O \bigg( {\alpha \over r^{ n
\ge 2}}
\bigg)$. Thus we arrive at the following expression:
\be
{d^2 \xi \over d r^2} + \bigg[ \omega^2 - m^2 + 
{4 M \omega^2 - 2 M m^2 \over r}
- {l(l + 1) \over r^2} \bigg] \xi = 0.
\ee
If one further assumes that the observer and the initial data are in
the region where $r \le {M \over (Mm)^2}$ and one shall be interested
in the intermediate asymptotic behavior of SI scalar field $r \le t
\le {M \over (Mm)^2}$, then we get
\be
{d^2 \xi \over d r^2} + \bigg[ \omega^2 - m^2
- {l(l + 1) \over r^2} \bigg] \xi = 0.
\label{wave}
\ee
The above approximation neglects the backscattering of the SI field
from asymptotically far regions. As in Ref.\cite{pir3} in our case of
the scalar field perturbations on dilaton black hole we get the
dependence on the field's parameter (the mass of the field).  These
perturbations do not depend on the spacetime parameters ($Q$ and $M$).
\par
The procedure of getting the solution of Eq.(\ref{wave}) is the same
as described in \cite{pir3} so we refer the reader to this work.  We
only write the final form of the $\psi$ dependence.  The intermediate
asymptotic behavior of the SI field at a fixed radius has the form
\cite{pir3}
\be
\psi \sim t^{- (l + {3 \over 2})} \sin mt,
\label{b1}
\ee
while the intermediate behavior of the considered field at dilaton
black hole outer horizon $H_{+}$ is dominated by
\be
\psi \sim v^{- (l + {3 \over 2})} \sin mt.
\label{b2}
\ee
They are dominated by an oscillatory power law tails.
\par
Recently the late-time behavior of the SI field in Schwarzschild
background was studied in Ref.\cite{ja1}. Following their footsteps we
shall write the wave Eq. (\ref{wave}) using the non-dimensional
variable $x = r/2M$. Then we reach to the analytically tractable
equation of the form
\be
{d^2 \psi \over d x^2} + {1 \over x (x - 1)} {d \psi \over d x} +
\bigg[
{4 M^2 \omega^2 x^2 \over (x - 1)^2} - {l(l + 1) \over x (x - 1)} -
{1 \over x^2 (x - 1)} + {4 m^2 M^2 x \over x - 1}
\bigg] \psi + O \bigg( {Q \over M} \bigg) = 0.
\label{jap}
\ee
The procedure of getting the solution to Eq.(\ref{jap}) was presented
in details in Ref.\cite{ja1}.  Here we quote the main conclusion,
namely the smaller value of $m M$ is the later the $t^{-5/6}$ tail
begins to dominate. It is true for the range of parameter $mt \gg {1
\over (m M)^2}$. For $mt \gg m M$ it was found that the larger value
of $m M$ is the later the $t^{-5/6}$ tail begins to dominate.  Of
course these conclusion are valid for small $Q$. In the next section
we investigate this problem numerically.

\section{NUMERICAL RESULTS}
\begin{figure}[ht]
\begin{center}
\leavevmode
\epsfxsize=220pt
\epsfbox{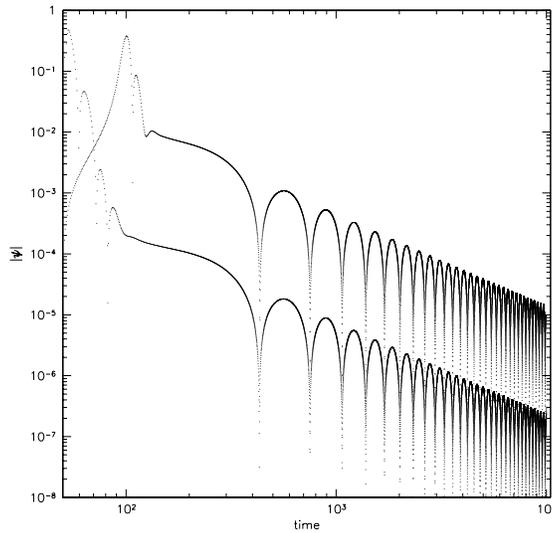}
\end{center}
\caption{Evolution of the field $\mid \psi \mid$ for $l=0$, $M=0.5$, $Q=0.45$
and $m=0.01$. Upper curve represents the field at fixed radius $y=50$
(approximation of the future timelike infinity $i_+$) as a function of
t. Lower curve represents the field at $u=10^4$ (approximation of the
black hole future horizon $H_+$) as a function of $v$. The power-law
exponents are $-1.51$ for both $i_+$ and $H_+$. The period of the
oscillations is $T=\pi/m \simeq 314.5$ to within $0.2\%$.}
\label{l0}
\end{figure}

We numerically analyzed Eq.~(\ref{mo}) using method described in
\cite{gun}. We transformed Eq.~(\ref{mo}) into $(u,v)$ coordinates
\be
4 \psi_{,uv} + V \psi = 0,
\ee
and solved it on uniformly spaced grid using explicit difference
scheme. As was pointed out previously the late time evolution of a
massive field is independent of the form of the initial data. To start
our calculation we used a Gaussian pulse of the form as follows:
\be
\psi(u=0,v) = A \exp \left ( - {(v-v_0)^2 \over \sigma^2} \right )
\ee
Linearity of the Eq.(\ref{mo}) leaves us a freedom to choose the
amplitude $A$, and for our purposes we used $A=1$. The rest of the
initial field profile parameters were $v_0=50$ and $\sigma=2$. The
black hole mass and charge are set equal to $M=0.5$ and $Q=0.45$,
respectively, and mass of the field is $m=0.01$.  We have studied the
behavior of the field $\psi$ on two hypersurfaces: the future timelike
infinity, $i_+$ (approximated in our calculations by the field at
fixed radius $y=50$), and the black hole future horizon $H_+$
(approximated by the field on the null surface $u=10^4$).  First, we
have examined $l=0$ modes. The results are presented in
Fig.~\ref{l0}. After initial period dominated by prompt contribution
and quasi-normal ringing the field establishes a definite oscillatory
pattern. The amplitudes of the field decrease according to power-law
with index $-1.51$ for both $i_+$ (top curve) and $H_+$.  These
exponents were calculated by fitting a straight line to the
oscillations extrema within the time range $2 \times 10^3 \div7 \times
10^3$ to avoid influence of the initial perturbation and the late time
tails. Analytically estimated from Eqs.(\ref{b1})-(\ref{b2}) value for
the index is $-1.5$ providing excellent agreement with numerical
results. The period of the oscillations (the distance between two
neighbor maxima) is $314 \pm 0.5$ which means less than $0.2\%$
discrepancy from analytically expected period $T = \pi/m\simeq 314.5$
\par
\begin{figure}[ht]
\begin{center}
\leavevmode
\epsfxsize=220pt
\epsfbox{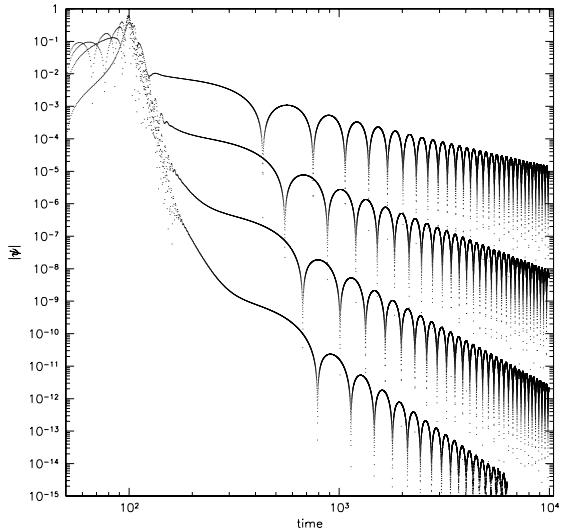}
\end{center}
\caption{Evolution of the field $|\psi|$ on the future timelike
infinity $i_+$ (approximated by $\psi(y=50,t)$) for different
multipoles $l=0,1,2$ and $3$. The power-law exponents are $-1.51$,
$-2.52$, $-3.53$, and $-4.51$, respectively. The period of the
oscillations is $314.5$ to within $0.5\%$ (for the worst case of $l=3$).}
\label{l0123}
\end{figure}

Secondly, we considered the dependence of the SI scalar field on the
multipole index $l$.  The results are presented in
Fig.~\ref{l0123}. The behavior of the field on the future timelike
infinity is calculated for $l=0,1,2$ and $3$. The numerical values of
the power-law exponents are $-1.51$, $-2.52$, $-3.53$, and $-4.51$,
respectively. According to Eq.(\ref{b1}) these exponents equal to
$-1.5$, $-2.5$, $-3.5$, and $-4.5$, thus again the agreement between
numerical calculations and analytical predictions is excellent. The
oscillations period is $T = \pi/m \simeq 315$ to within $0.5\%$, in
agreement with the predicted value.
\par
\begin{figure}[ht]
\begin{center}
\leavevmode
\epsfxsize=220pt
\epsfbox{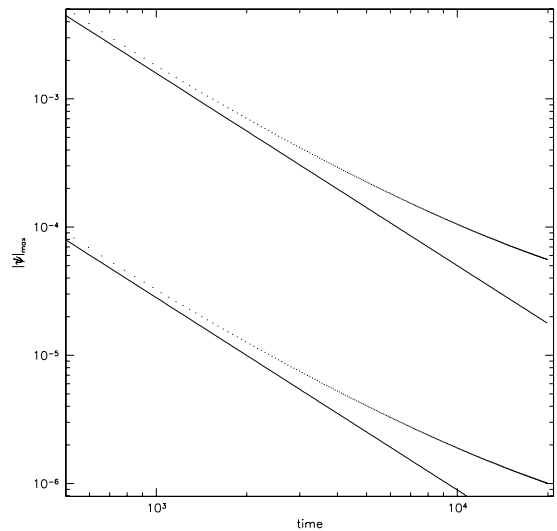}
\end{center}
\caption{Late-time behavior of the field $\mid \psi \mid_{max}$ on
dilaton black hole. The mass of the field $m = 0.05$, the upper curve
represents the field at future timelike infinity $(y = 50)$ as a
function of time $t$. The lower curve along the black hole horizon as
a function of $v$. The period of oscillations is $63.0$ to within
$0.8\%$. The thin solid lines have slopes equal to $-1.5$.  }
\label{late}
\end{figure}

Finally, we have checked the late-time behavior of the field. In
Fig.~\ref{late} we present the field on the future timelike infinity
$i_+$ and on the black hole horizon $H_+$ as a function of time for
$l=0$ mode (for presentation purposes we plot only maxima of the
oscillations).  As in \cite{mm} our numerical scheme reveals the fact
that after the intermediate phase of the power-law decay the amplitude
of the field decreases slower than any power-law.  In the very late
approximation the change of the tail will be result of the dominant
backscattering due to the spacetime curvature.  This effect is beyond
the flat space approximation as was remarked in \cite{ja}.
Further, we fix mass and charge of dilaton black hole and change the
mass of the SI scalar fields. We have noted the relation between mass
$m$ and the another pattern of decay; namely, the smaller $m$ is the
later begins to dominate this pattern. We remark that analogical
relation was obtained from analytical considerations when the charge
of dilaton black hole was assumed to be small.  In our numerical
simulations we get rid of this assumption.
\begin{figure}[ht]
\begin{center}
\leavevmode
\epsfxsize=220pt
\epsfbox{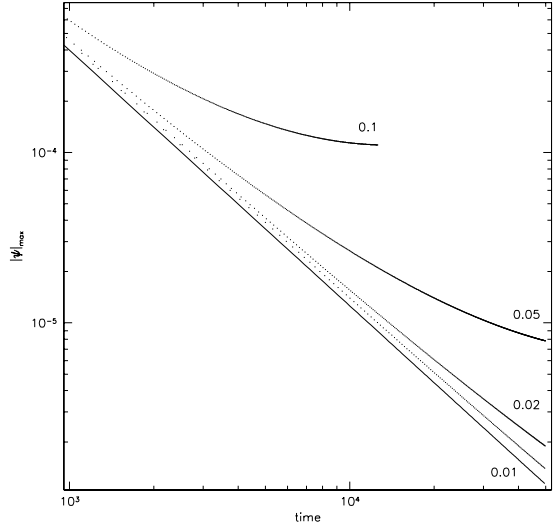}
\end{center}
\caption{Late-time behavior of the SI scalar field $\mid \psi \mid_{max}$ 
(maxima of the oscillations) at future timelike infinity for different
masses $m$ as a function of time $t$.  The thin solid line has slope
equals to $-1.5$, mass of dilaton black hole $M = 0.5$ and charge $Q =
0.45$.}
\label{mmm}
\end{figure}

\section{Conclusions}
In our work we have investigated the intermediate asymptotic behavior
of SI scalar field in the background of dilaton black hole being the
spherical symmetric solution of the so-called low-energy string theory
\cite{dbh}. As was revealed the late-time behavior of the charged
massless perturbations depended on the spacetime parameters $M, Q$,
while the intermediate asymptotic behavior of SI field depended only
on the field's mass $m$. The numerical scheme has shown that at the
very late-time the inverse power-law decay is replaced by another kind
of the decay process, slower than any power-law.  From our numerical
calculations we have established that if one fixed the mass and charge
of dilaton black hole $M$ we get the relation: the smaller SI field
mass is, the later begins the process.  This very late-time behavior
needs further analytic studies. The first step towards it was done in
\cite{ja,ja1} studying nearly extreme RN and Schwarzschild solutions.
We have found that in the case of small $Q$ comparing to $M$, our wave
equation reduce to the form studied in \cite{ja1}. Thus, it can be
deduced that the asymptotic late-time behavior of SI scalar field is
of the form of the oscillatory tail with the decay rate $t^{-5/6}$,
which is identical with that in the nearly extreme RN or Schwarzschild
background.
\par
Another problem for further investigations is the problem of late-time
behavior of massless and SI scalar fields in the spacetime of extremal
dilaton black hole.  We hope to return to this problem elsewhere.



\end{document}